\documentstyle[11pt,epsfig,a4]{article}
\begin{document}

\def\Journal#1#2#3#4{{#1} {\bf #2}, #3 (#4)}

\def\etal{{\it et\ al.}}
\def\NCA{\em Nuovo Cim.}
\def\NIM{\em Nucl. Instrum. Methods}
\def\NIMA{{\em Nucl. Instrum. Methods} A}
\def\NPB{{\em Nucl. Phys.} B}
\def\PLB{{\em Phys. Lett.}  B}
\def\PRL{\em Phys. Rev. Lett.}
\def\PRC{{\em Phys. Rev.} C}
\def\PRD{{\em Phys. Rev.} D}
\def\ZPC{{\em Z. Phys.} C}
\def\ASP{{\em Astrop. Phys.}}
\def\JETP{{\em JETP Lett.\ }}

\def\numunue{\nu_\mu\rightarrow\nu_e}
\def\numunutau{\nu_\mu\rightarrow\nu_\tau}
\def\nue{\nu_e}
\def\nuebar{\bar\nu_e}
\def\numubar{\bar\nu_\mu}
\def\numubarnuebar{\bar\nu_\mu\rightarrow\bar\nu_e}
\def\dm2{\Delta m^2}
\def\numu{\nu_\mu}
\def\nutau{\nu_\tau}
\def\ra{\rightarrow}

\vspace*{-2cm}
\begin{center}

\date{ 26 May 1999}
\vskip 1.5cm

\Large{\bf Three-family oscillations using neutrinos \\
from muon beams at very long baseline}
\vskip .25in
\normalsize{M. Campanelli$^1$, A. Bueno$^1$, A. Rubbia$^{1,2}$}\\
{\it $~^1$ Institut f\"{u}r Teilchenphysik, ETHZ, CH-8093 Z\"{u}rich, Switzerland}\\
{\it $~^2$ CERN, Geneva, Switzerland}
\end{center}
\vskip 0.5cm

\abstract{
The planned LBL experiments will be able to prove the hypothesis
of flavor oscillation between muon and tau neutrinos.
We explore the possibility of a second generation long
baseline experiment at very long baselines, i.e. $L$ in the
range $6500-8800\ \rm km$. This distance requires intense neutrino
beams that could be available from very intense muon beams as
those needed for $\mu$ colliders.
Such baselines allow to study neutrino oscillations with
$E/L \approx 2\times 10^{-3}\ \rm eV^2$ with neutrinos of energy
$E_\nu \approx 20\ \rm GeV$, i.e. above tau threshold.
Moreover, matter effects inside the Earth could lead to observable
effects in $\nu_e\to\nu_\mu$ oscillations. These
effects are interchanged between neutrinos and antineutrinos, 
and therefore they can be tested by comparing the
oscillated spectra obtained running the storage ring 
with positive and negative muons.}

\section{Introduction}
The use of neutrinos from muon decays has several advantages:

\begin{itemize}
\item easy prediction of the neutrino fluxes and flavors
(since no hadronic processes involved) when the muon polarization
is known;
\item flexibility in the choice of the beam energy, since precisely
determined by the muon storage ring energy.
\end{itemize}
These beams will contain neutrinos of the electron and muon flavors
in same quantity, i.e. $\mu^-\to e^-\nu_\mu \bar{\nu}_e$ or
$\mu^+\to e^+\nu_e \bar{\nu}_\mu$, a feature that distinguishes
them from traditional neutrino beams where $\nu_\mu$ dominates
(since $\pi\rightarrow e\nu_e$ is suppressed).

This well-defined flavor 
composition can be exploited using a detector with charge identification
capabilities to search for neutrino oscillations.
In Ref.~\cite{sbm}, we have studied a possible 
$\nu_\mu\to\nu_e$ oscillation search to study the LSND claim 
with a detector placed at a medium baseline.

Perhaps the most interesting option is the fact that
muon beams will be located next to muon colliders which necessarily
require very intense proton sources. This opens the possibility to
obtain very intense neutrino beams to perform long and very long baseline
oscillation experiments.

In Ref.~\cite{lbm}, we have considered a $\nu_\mu$ disappearance experiment 
with a baseline of 730~km. Since the neutrino beams contain both electron
and muon flavors, the disappearance of muon neutrino can be performed
by directly comparing electron and muon events. This method is self
normalizing, i.e. without the need of a near 
detector to predict the original flux. 

Therefore, the very attractive feature for the future, rather than to 
think of experiments
on currently accessible baselines (i.e. the FNAL-NUMI to Soudan and
the CERN-NGS to GranSasso) is to
imagine a second generation of long baseline experiments with
very long baseline of the order of 6500-8800~km, made possible
by the intense neutrino sources.

\section{Oscillation searches at VLBL}

The distance L between neutrino production and detection is a key element in
oscillation studies. Since the oscillation probability varies like 
$$\sin^2(1.27\Delta m^2(eV^2) L(km)/E_\nu(GeV)),$$ 
having a longer length means the possibility to explore 
the same parameter-space region with neutrinos of higher energy.

In addition, since these beams will travel through Earth for thousands
of kilometers, matter enhancement in neutrino oscillation \cite{msw}
can become important, as recently pointed out in Ref.~\cite{lipari}.
We perform oscillation searches that can be categorized as follows:
\begin{itemize}

\item {\bf $\mu$ disappearance}:
similar to what we described in Ref.~\cite{lbm}; but 
improved due to the larger L/E ratio, to observe the 
``dip'' in the $\nu_\mu$ energy distribution, even for the lowest 
$\Delta m^2$ suggested by the atmospheric neutrino effect.

\item {\bf $e$ and $\mu$ appearances}:
study of matter effects in $\nu_e\to\nu_\mu$ oscillations\par
Travel through the Earth would enhance the
transition probability for neutrinos (antineutrinos) if $\Delta m^2>0$ 
($\Delta m^2<0$). Running the muon storage ring alternatively with
negative and positive muons would directly reveal this effect and
determine the sign of $\dm2$.

\item {\bf $\tau$ appearance}: the production of $\tau$ leptons can be determined looking for an excess of
hadronic events due to hadronic $\tau$ decays.

\end{itemize}
We do not consider the possibility of direct tau appearance by identification of the
$\tau$ lepton using kinematical or topological criteria, and postpone it to
further studies.

\section{Possible very long baselines}
We assume that a multikiloton detector will be located in Europe
at the Laboratori Nazionali del GranSasso (LNGS, Italy). 

The ideal detector would be a 10 kiloton (fiducial) SuperICARUS
as described in Ref.~\cite{sica}. In addition to providing excellent spatial
resolution like that of bubble-chamber, SuperICARUS, unlike most neutrino detectors, 
allows for excellent
electron identification and measurement. High energy muons, exiting the detector,
can also be measured from multiple scattering with a resolution $\Delta p/p \approx 20\%$.
These unique lepton capabilities are in a sense matched to the beams from muon decays,
which provide equal amounts of electron and muon neutrinos.

We also assume that a ``sign-determining'' muon spectrometer 
is installed behind the target, to
allow the identification of the leading muon charge (``$\mu^-$'', ``$\mu^+$'' samples). 
On the other hand, the leading
electron charge cannot be identified (``$e$'' sample).
This is due to the fact that electrons will produce a shower 
soon after production before being sufficiently bent in a possible 
magnetic field to be measured.\par
\begin{figure}[p]
  \begin{center}
   \includegraphics[width=15cm]{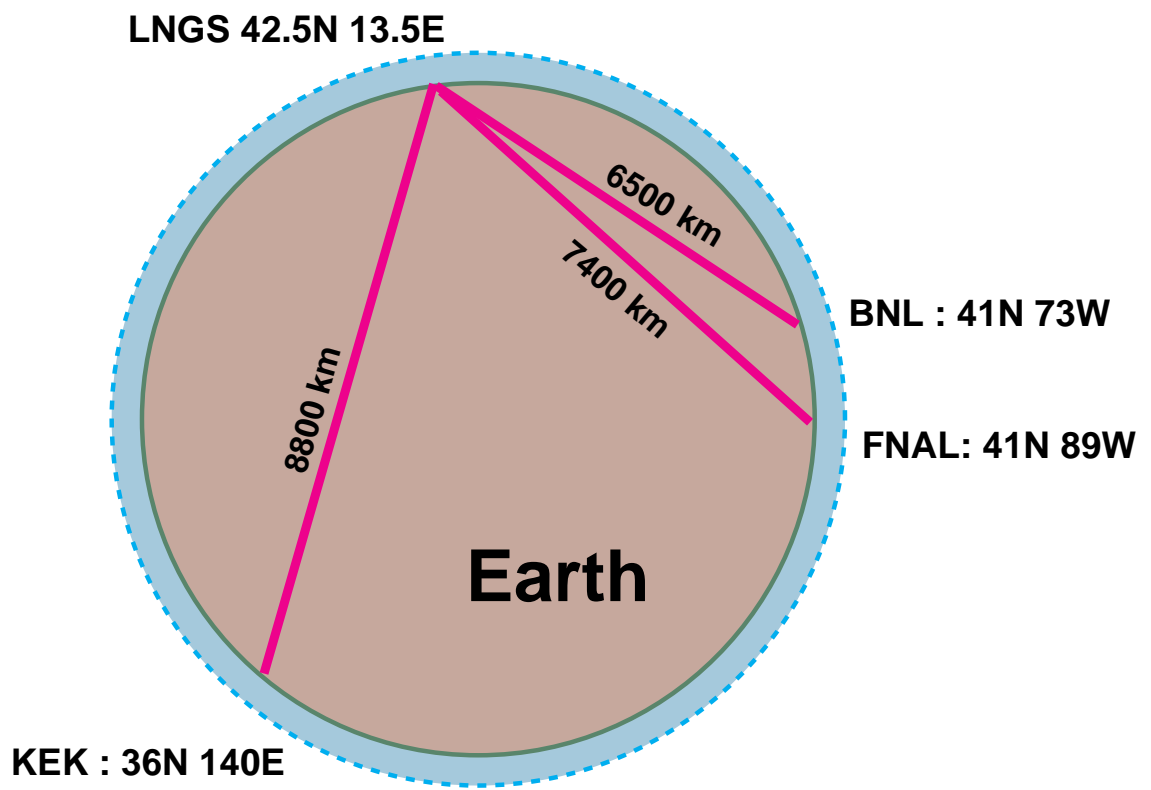}
  \end{center}
\caption{Possible very long baselines across the Earth (seen from above the North pole).}
\label{fig:baseline}
\end{figure}

The neutrino source could be located in different laboratories around the world,
in the American or Asian continent. For example, we quote BNL-GranSasso,
FNAL-GranSasso and KEK-GranSasso.
The BNL-LNGS has a baseline of $L=6500\ \rm km$ (see Figure~\ref{fig:baseline}).
The beam goes to a maximum depth of 900~km and arrives at GS with
an angle with respect to horizontal of about $30^o$. We estimate
the average density of the Earth\cite{bahc} for this baseline
to be $\rho=3.6\ \rm g/cm^3$.
Other baselines are
FNAL-GranSasso ($L=7400\ \rm km$, max. depth 1200 km, average
density $\rho=3.7\ \rm g/cm^3$, beam angle $36^o$) or KEK-GranSasso 
($L=8800\ \rm km$,
max. depth 1800 km, average density $\rho=4.0\ \rm g/cm^3$, beam
angle $44^o$). There is not much differences in $L$ between
the various baselines.

Our attention was recently drawn to the upgrade of the AGS accelerator
in BNL\cite{palmer}. A total
of $6.6\times 10^{21}$ protons on target per year could be produced.
When used as proton driver of a muon accumulator with a muon yield per proton of 15\%, 
we obtain $6.6\times 10^{21}\times 0.15 = 9.9\times 10^{20}$ muons/year.
The machine would be operated for four years, 
alternating runs with positive and
negative muons and we consider that, given the shape of the muon storage ring,
half of the muons will decay
in the direction towards the GranSasso laboratory. We then use
as integrated intensity a total of
\[ N(\mu^+)+N(\mu^-)= \frac{1}{2}\times6.6\times 10^{21}\times 0.15\times 4\approx 2\times 10^{21}\mu\rm 's \]
decay in the right direction producing neutrinos.

\section{Event rates for the BNL/AGS-GS baseline}
Neutrinos travel the distance between the BNL and GS laboratories
and are then detected in the 10 kton (fiducial) detector. 

The neutrino event rate will grow as $E_\mu^3$ where $E_\mu$ is the
energy of the muon storage ring. We are interested in a neutrino
beam with energy of about $20\rm \ GeV$. We list
in Table~\ref{tab:ratenonosc} the neutrino event rates in case
of no oscillation as a function of the muon storage ring energy,
for $10^{21}$ muons decays of a given charge and 10 kton target.
Neutrino interactions have been divided into charged
current (CC) and neutral current (NC) interactions.
The CC events are split into inelastic
scattering (DIS) and quasi-elastic (QE) interactions.

\begin{table}[tbh]
\begin{center}
\begin{tabular}{|c||c|c|c||c|c|c|}\hline
& \multicolumn{3}{c||}{$10^{21}\mu^-$ decays} & \multicolumn{3}{c|}{$10^{21}\mu^+$ decays} \\
$E_\mu$ (GeV)&CC($\bar{\nu}_e$)&CC($\nu_\mu$)&NC&CC($\nu_e$)&CC($\bar{\nu}_\mu$)&NC\\ \hline
  10&  426& 1152&  196& 1016&  488&  173\\
  15& 1414& 3751& 1129& 3283& 1624& 1005\\
  20& 3313& 8712& 3542& 7588& 3820& 3172\\
  25& 6412&16746& 8221&14568& 7401& 7419\\
  30&11010&28576&16008&24850&12710&14524\\ \hline

\end{tabular}
\end{center}
\caption{The total number of neutrinos detected in a 10 kton (fiducial) detector 
for a baseline $L=6500\ \rm km$ and a total number of $10^{21}$ muons decays.}
\label{tab:ratenonosc}
\end{table}

\section{Neutrino oscillation scenarios}
In the preceding section, we have seen that the expected neutrino event
rates are comfortable and open possibilities to study various neutrino
oscillations scenarios with a baseline of 
$$\frac{E}{L} \approx \frac{20\ \rm GeV}{6500\ \rm km} \approx 2.5\times 10^{-3}\ \rm eV^2.$$

While the atmospheric neutrino evidence is compatible with maximal mixing
between $\nu_\mu$ and $\nu_\tau$, it is attractive to think that there is also
a smaller mixing between the first and third family.
We therefore consider a three-family scenario. The mixing between the different neutrino
flavors is determined by two mass differences and a unitary matrix describing
the mixing between flavor and mass eigenstates. 

We base our computation on the formalism of Ref.~\cite{lipari} and
make the approximation that only a mass scale is relevant, i.e.
assuming that the differences $|m_3^2-m_2^2|\gg
|m_2^2-m_1^2|$ (this approximation being supported by 
current experimental results on solar neutrinos).
The mass eigenstate $m_1$ is defined orthogonal to the electron flavor state.
The mixing matrix takes a form that only depends on the two
mixing angles $\theta$ and $\phi$:
\begin{equation}
U=\left(
\begin{tabular}{ccc}
0&$\cos\theta$&$\sin\theta$\\
$\cos\phi$&$-\sin\theta\sin\phi$&$\cos\theta\sin\phi$\\
$-\sin\phi$&$-\sin\theta\cos\phi$&$\cos\theta\cos\phi$\\
\end{tabular}\right)
\end{equation}

\par
The three-family oscillation is therefore described by only three parameters:
the mass difference between the second and third neutrino 
$\Delta m^2\equiv\Delta m_{23}^2$ and the two mixing angles $\theta$ and $\phi$.
\par
The long travel inside the Earth gives rise to matter effects,
resulting in a modification of the oscillation.
The mixing angle in matter $\theta_m$ takes a value given by the expression
\[\sin^2 2\theta_m(x)=\frac{\sin^2 2\theta}{\sin^2 2\theta+ (x-\cos 2\theta)^2}\]
and
\[x=\frac{2V E_\nu}{\Delta m^2}=\pm\frac{2\sqrt{2} G_F n_e E_\nu}{\Delta m^2},\]
where the plus sign applies to $\nu$'s and the minus to $\bar{\nu}$'s
and $n_e$ is the electron density of the medium.
In the limit of no matter effects $x\to0$ and the mixing angle $\theta_m\to
\theta$.
Given the distance between the two laboratories, the  beam reaches the maximal
depth of about 900 km, where the matter density of the earth is larger than 
that close to the surface (we derive this information from Ref.~\cite{bahc}).
An average matter density of 3.6 $g/cm^3$ was considered.

\section{Oscillated neutrino fluxes}
We assume that the mixing is maximal between the second and the third neutrino
family ($\sin^2\phi=0.5$) and small ($\sin^2\theta=0.025$) between the first
two. We will consider the three values of $\Delta m^2=10^{-3}, 
3\times 10^{-3}$ and $10^{-2}\ \rm eV^2$.
\par
In the case of a neutrino beam from negative muons, four oscillation processes will occur:
\begin{itemize}
\item $\nu_\mu\to\nu_e$
\item $\nu_\mu\to\nu_\tau$
\item $\bar{\nu}_e\to\bar{\nu}_\mu$
\item $\bar{\nu}_e\to\bar{\nu}_\tau$
\end{itemize}
while for the case of positive muons the charge-conjugate processes 
occur.
\par
\begin{figure}[p]
  \begin{center}
   \includegraphics[width=\linewidth]{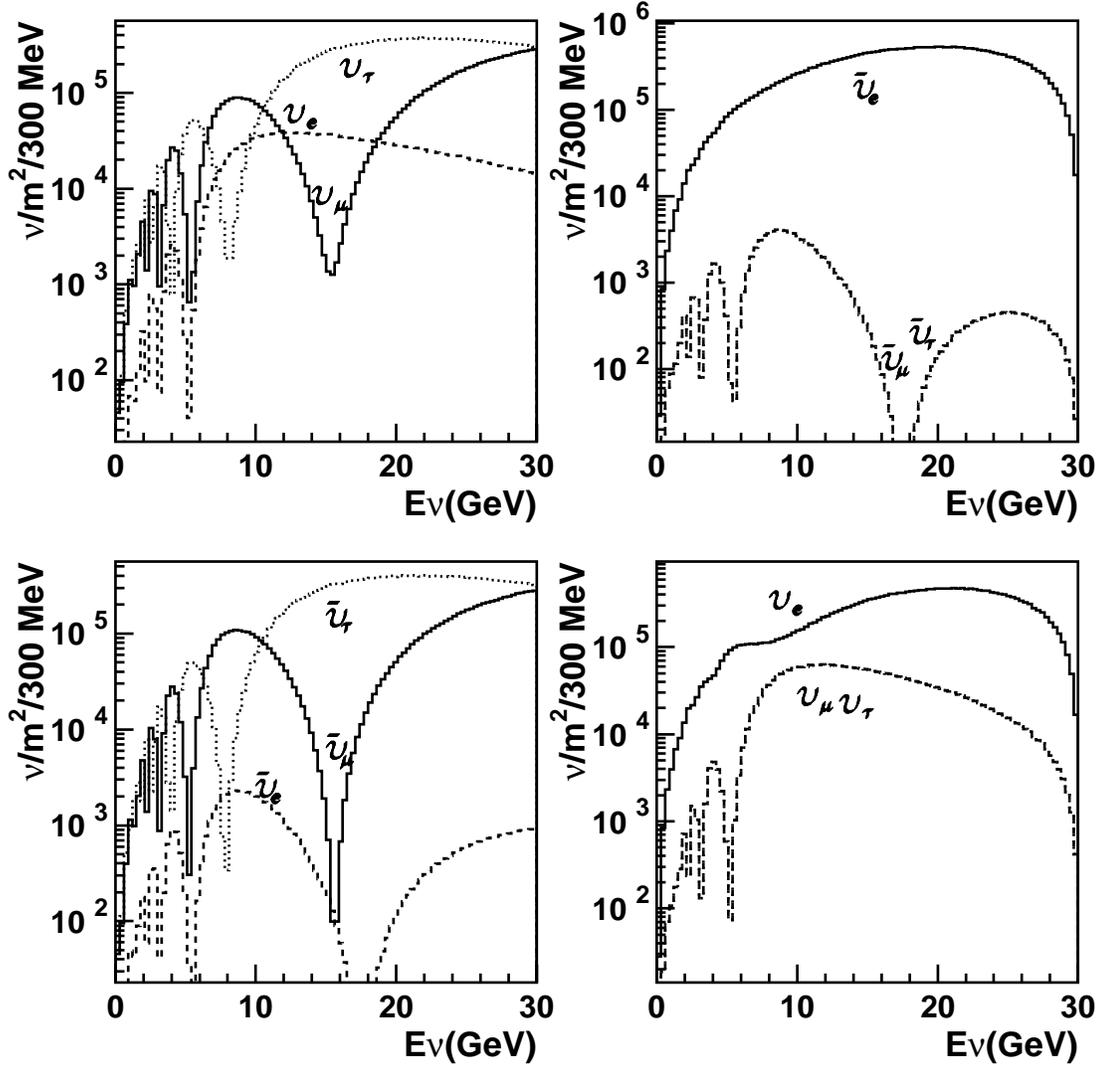}
  \end{center}
\caption{
The oscillated neutrino fluxes reaching the detector per $m^2$ normalized
to $10^{21}$ muon decays. The oscillation probability is given for three-family 
mixing with $\Delta m^2=3\times 10^{-3}eV^2, \sin^2\phi=0.5,
\sin^2\theta=0.025$. The two upper plots refer to decays of 30 GeV $\mu^-$, 
the two lower to decays of $\mu^+$. The matter enhancement(suppression) is
clearly visible for the $\nue$($\bar\nue$) cases. The $\numu$($\bar\numu$)
fluxes are largely suppressed with a spectacular ``hole'' in the spectrum
due to maximal oscillation to $\nutau$. The $\nue\ra\numu$ and $\nue\ra\nutau$
($\bar\nue\ra\bar\numu$ and $\bar\nue\ra\bar\nutau$)
contributions are also visible.}
\label{fig:fluxes}
\end{figure}

The neutrino fluxes as a function of energy for the different flavors
in the case of three-family oscillations are shown in Figure 
\ref{fig:fluxes} for a $30\ \rm GeV$ muon beam and
$\dm2 = 3\times 10^{-3}\ \rm eV^2$.

\section{Event classification}
We classify the observed events in four classes:
\begin{itemize}
\item[a)] events with electrons or positrons (no electron charge measured),
\item[b)] events with muons of the same sign of those circulating
in the storage ring,
\item[c)] events with muons of opposite sign,
\item[d)] events without leptons.
\end{itemize}

No direct $\tau$ identification is performed, so, according to their decay,
$\tau$ events are seen either in the electron ($\tau\to e$ decays),
muon ($\tau\to \mu$) or neutral current ($\tau\to h$) sample.\par

\begin{figure}[p]
  \begin{center}
   \includegraphics[width=\linewidth]{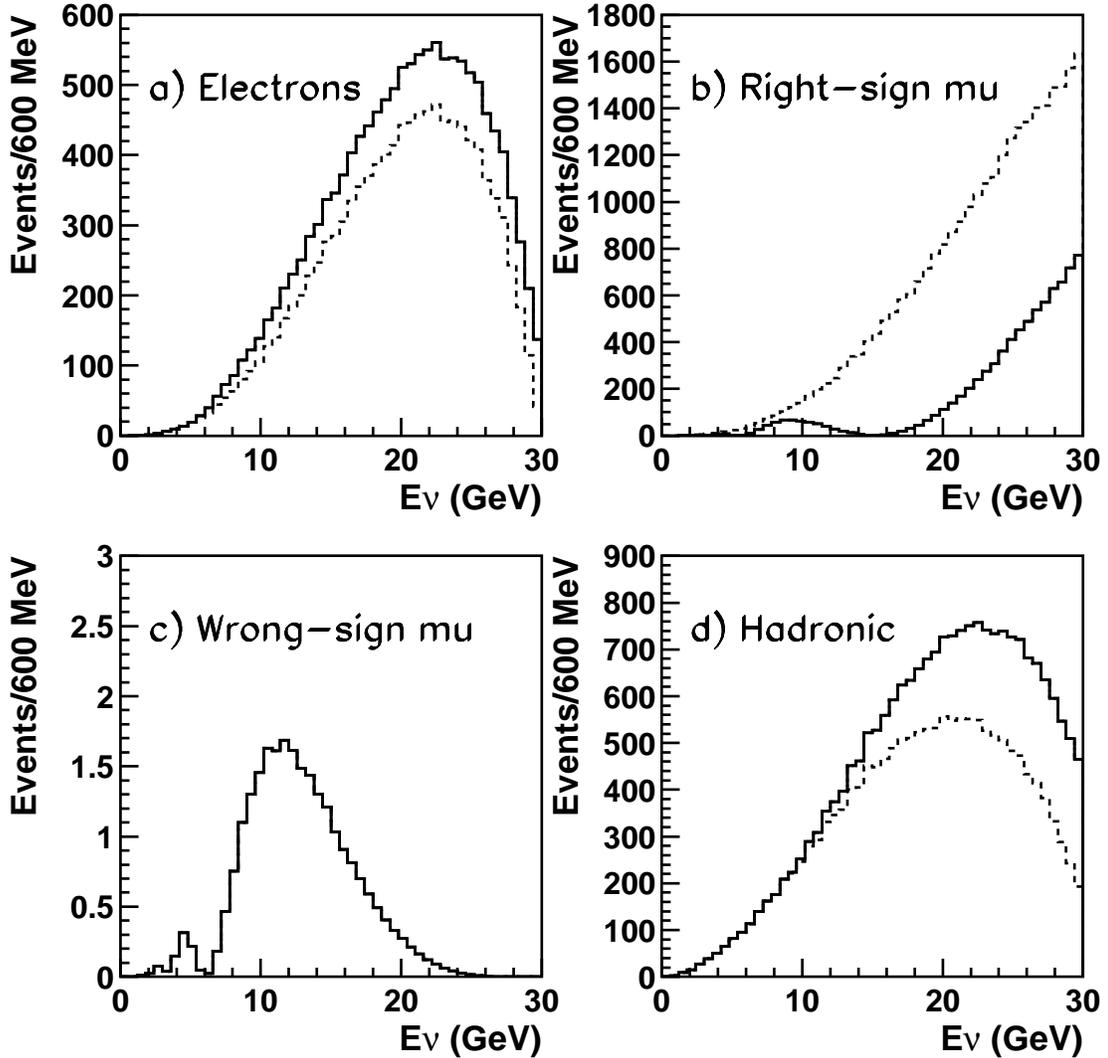}
  \end{center}
\caption{Predicted event energy spectrum of
the four classes of events a) with leading electron or positron,
b) with leading right-sign muons, c) with leading opposite
sign muons d) events with no leading leptons.
Full line: spectra with oscillations; dashed line: spectra
without oscillations. 
The plots are for $10^{21} \mu^-$ decays. The muon energy is
30~GeV and the target mass is 10 kton. The spectacular
disappearance of right-sign muon is visible in plot b).
The suppression due to matter effects
is clearly visible in plot c) which contain the opposite-sign
muon events. The apperance of tau can be directly observed
as an excess of neutral current like events (plot d).
}
\label{fig:mu-spec}
\end{figure}

\begin{figure}[p]
  \begin{center}
   \includegraphics[width=\linewidth]{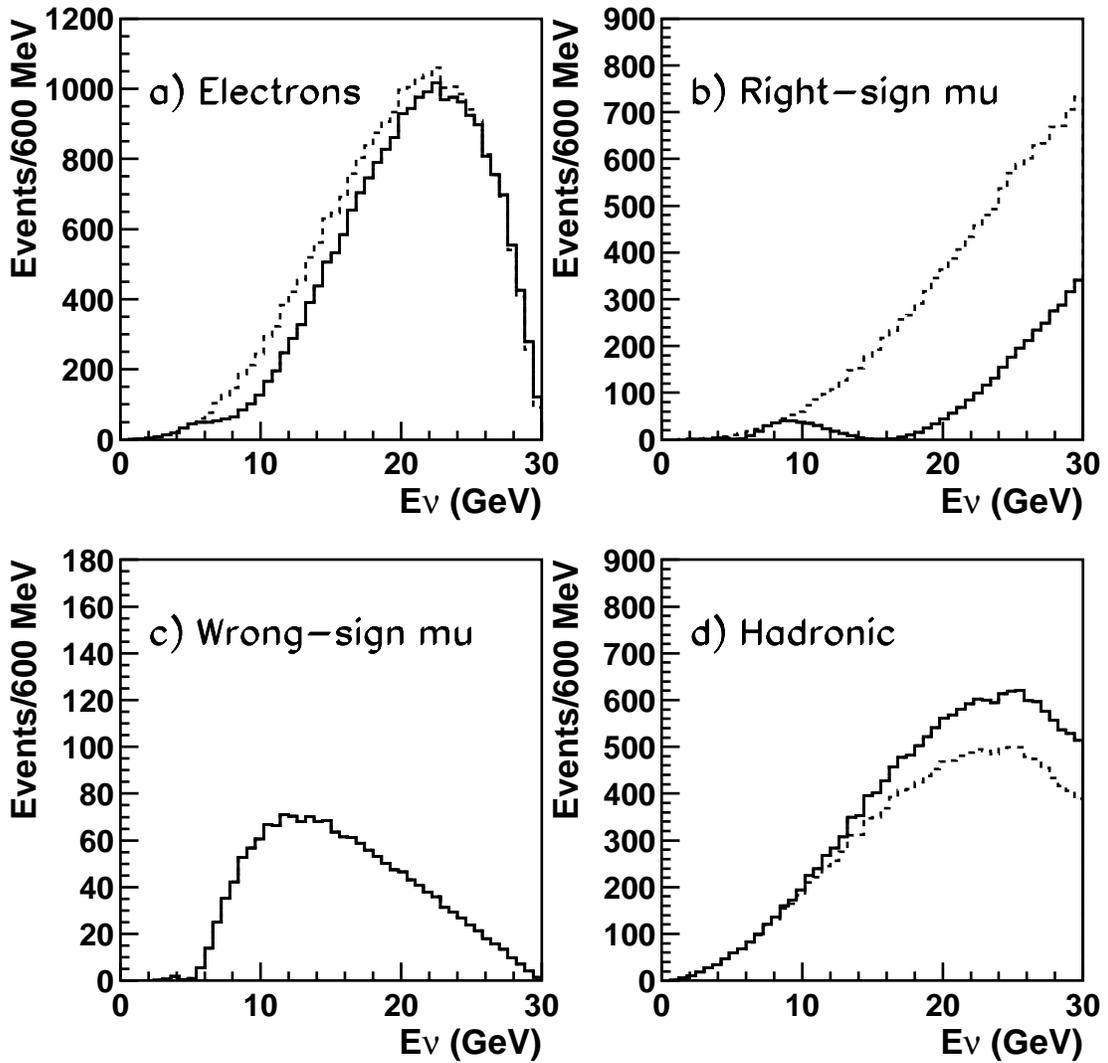}
  \end{center}
\caption{Same as previous plot, but 30 GeV $\mu^+$ decays.
The sign of the muon charges are also exchanged, i.e. plot b refers to
events with leading $\mu^+$, plot c to events with leading $\mu^-$.
In this case, the enhancement due to matter effects is clearly
visible.}
\label{fig:mu+spec}
\end{figure}

Figure \ref{fig:mu-spec} shows the energy spectra of the four classes 
for a 30 GeV $\mu^-$ beam.
The dotted and full lines refer to the 
predicted distributions without oscillations, and with oscillations
with $\Delta m^2=3\times 10^{-3} eV^2$. 
In Figure \ref{fig:mu+spec} the same distributions are shown for 30 GeV
$\mu^+$ beams.
\par
Let us consider in more detail the
processes playing a role for the different classes of observed events. 
In the following, 
every process is listed with the relevant probability, where the notation 
$P_{\nu_\alpha\to\nu_\beta}$ is used to identify the probability of 
neutrinos of flavor $\alpha$ to oscillate into neutrinos of flavor 
$\beta$. These probabilities (and the text,
unless otherwise specified) refer to the
case of negative muons in the storage ring; the charge-conjugate must be
considered for the case of positive muons.

\subsection{Events with electrons}
In these events, a leading electron or positron is identified in the detector.
Such events could be produced by the charged-current interactions of the
following neutrinos:
\begin{enumerate}
\item unoscillated $\bar{\nue}$ neutrinos from the beam\par
$(1-P_{\bar{\nu}_e\to\bar{\nu}_\mu}-P_{\bar{\nu}_e\to\bar{\nu}_\tau})
\times\Phi_{\bar{\nu}_e}$
\item muon neutrinos oscillated into electron neutrinos\par
$(P_{\nu_\mu\to\nu_e})\times \Phi_{\nu_\mu}$
\item tau neutrinos derived from oscillations followed by a $\tau\to e$ decay\par
$(P_{\nu_\mu\to\nu_\tau}\times BR(\tau^-\to e))\Phi_{\nu_\mu}+
(P_{\bar{\nu}_e\to\bar{\nu}_\tau}\times BR(\tau^+\to e))\Phi_{\bar{\nu}_e}$
\end{enumerate}
In case of $\mu^-$ beams, process 1) will deplete the number
of leading electron events from the beam, while process 2) will increase
the number of such events, as they come from oscillation of muon neutrinos.
Since process 2) involves neutrinos, it is matter-enhanced, and the net
effect will be an increase of the number of observed events having a 
leading electron with respect to the case of no oscillation, as can
be seen comparing the Figures in Table \ref{tab:ele-} with the 
non-oscillated case (Table \ref{tab:ratenonosc}).\par
In the case of $\mu^+$ beams, the balance of the two competing processes
is reversed. Now process 1) involves neutrinos, so the depletion of
electron events from the beam is matter-enhanced, while process 2) is suppressed.
The net effect is a smaller number
of observed events with leading electrons with respect to the expectations,
as can be seen comparing Tables \ref{tab:ele+} and \ref{tab:ratenonosc}.
\par
\begin{table}[tbh]
\begin{center}
\begin{tabular}{|c|c|c|c|c|c|}\hline
$E_\mu$ &$\Delta m^2 (eV^2)$&  unoscillated beam &    $\nu_\mu\to\nu_e$ &   $\tau$ decay\\ 
(GeV) & & $(1-P_{\bar{\nu}_e\to\bar{\nu}_x})
\times\Phi_{\bar{\nu}_e}$ &  $(P_{\nu_\mu\to\nu_e})\times \Phi_{\nu_\mu}$   &
$((P_{\nu_\mu\to\nu_\tau})\Phi_{\nu_\mu}+(P_{\bar{\nu}_e\to\bar{\nu}_\tau})\Phi_{\bar{\nu}_e})$
\\ & & & &  $\times BR(\tau\to e)$\\
\hline
  10&                  &  424.&   56.&   17\\
  20&$1.\times 10^{-3}$& 3309.&   76.&  118\\
  30&                  &11011.&   85.&  271\\ \hline
  10&                  &  418&  252&    6\\
  20&$3.\times 10^{-3}$& 3268&  977&  380\\
  30&                  &10967& 1366& 1423\\ \hline
  10&                  &  411&   40&   11\\
  20&$1.\times 10^{-2}$& 3213&  449&  314\\
  30&                  &10820& 5805&  574\\ \hline
\end{tabular}
\end{center}
\caption{Different contributions to leading electron events from $\mu^-$ 
beams.}
\label{tab:ele-}
\end{table}

\begin{table}[tbh]
\begin{center}
\begin{tabular}{|c|c|c|c|c|}\hline
$E_\mu$ &$\Delta m^2 (eV^2)$&  unoscillated beam &    $\bar\nu_\mu\to\bar\nu_e$ &   $\tau$ decay\\ 
(GeV) & & $(1-P_{\nu_e\to\nu_x})
\times\Phi_{{\nu}_e}$ &  $(P_{\bar\nu_\mu\to\bar\nu_e})\times \Phi_{\bar\nu_\mu}$   &
$((P_{\bar\nu_\mu\to\bar\nu_\tau})\Phi_{\bar\nu_\mu}+(P_{{\nu_e}\to\nu_\tau})\Phi_{\nu_e})$
\\ & & & &  $\times BR(\tau\to e)$\\\hline
  10&                  &  874.&    1.&   10\\
  20&$1.\times 10^{-3}$& 7377.&    1.&   58\\
  30&                  &24624.&    3.&  128\\ \hline
  10&                  &  644&    5&    7\\
  20&$3.\times 10^{-3}$& 5413&   17&  233\\
  30&                  &21457&   15&  761\\ \hline
  10&                  &  941&   10&    6\\
  20&$1.\times 10^{-2}$& 6929&   45&  161\\
  30&                  &17208&  120&  521\\ \hline
\end{tabular}
\end{center}
\caption{Different contributions to leading electron events from $\mu^+$ 
beams.}
\label{tab:ele+}
\end{table}
\par

\subsubsection{Quasi-elastic events}
The electron charge is not measured. 
In order to identify neutrino or antineutrino interactions without the 
need of a measurement of the electron charge, quasi-elastic interactions 
can however be used.
In particular, 
\[\nu_e+n\to e^-+p\]
 events are quite cleanly 
distinguishable from DIS processes, due to the presence of a proton, seen 
in the detector as a highly ionizing track. The same does not hold for 
antineutrino interactions
\[\bar{\nu}_e+p\to e^++n\] 
since the neutrons do not produce any track at the vertex.

The presence of
$e+p$ final state in a beam from $\mu^-$ is therefore a clear signature for 
$\nu_\mu\to\nu_e$ oscillation.  
The total rate of $e^- p$ events produced in quasi-elastic interactions 
from a beam of negative muons is listed in Table \ref{tab:qe}. 
The number of quasi-elastic events is non negligible, even after 
taking into account the
detection efficiency due to a necessary proton momentum threshold for
detection and rejection of nuclear spectators protons.
\par

\begin{table}[tbh]
\begin{center}
\begin{tabular}{|c|c|c|c|c|c|c|}\hline
$E_\mu$ (GeV)&$\Delta m^2 (eV^2)$&$e^- p$ events (QE)&$e^+ n$ events (QE)\\ 
& & $(P_{\nu_\mu\to\nu_e})\times \Phi_{\nu_\mu}$ & $(1-P_{\bar{\nu}_e\to\bar{\nu}_x})\times\Phi_{\bar{\nu}_e}$ \\
\hline
  10&                  &   7& 58\\
  20&$1.\times 10^{-3}$&   7&258\\
  30&                  &   6&598\\ \hline
  10&                  &  21& 57\\
  20&$3.\times 10^{-3}$&  55&254\\
  30&                  &  61&595\\ \hline
  10&                  &   4& 56\\
  20&$1.\times 10^{-2}$&  21&250\\
  30&                  & 169&587\\ \hline
\end{tabular}
\end{center}
\caption{Number of quasi-elastic interactions ($e^-p$ final states)
from neutrinos produced in $\mu^-$ beams}
\label{tab:qe}
\end{table}

\subsection{Events with right-sign muons}
The muon charge is measured by a spectrometer placed behind the detector.
Muons of the same
sign of those decaying in the storage ring are produced by:\begin{enumerate}
\item muon neutrinos coming from the beam\par
$(1-P_{\nu_\mu\to\nu_e}-P_{\nu_\mu\to\nu_\tau})\times\Phi_{\nu_\mu}$ 
\item $\tau$ decays, where the $\tau$ is coming from $\nu_\mu\to\nu_\tau$
or $\bar{\nu}_e\to\bar{\nu}_\tau$ oscillations.\par
$P_{\nu_\mu\to\nu_\tau}\times BR(\tau\to\mu)\times\Phi_{\nu\mu}$\par
\end{enumerate}
Since in this case we measure the muon sign, $\nu_\mu$ oscillating
into $\nu_e$ are not compensated by $\bar{\nu}_e$ oscillations into 
$\bar{\nu}_\mu$, as they produce muons of opposite charge (see following
subsection). A small compensation to the total number of right-sign muons
comes from $\nu_\mu\to\nu_\tau$ oscillations where the $\tau$ decays into
a muon. For both $\mu^-$ and $\mu^+$ events, the net effect of neutrino
oscillations is a disappearance of right-sign leading mung-muon events.

\subsection{Events with wrong-sign muons}
Opposite-sign muons can only be produced by neutrino oscillations, since
there is no component in the beam that could account for them. 
These events are coming from oscillations of the electron component of 
the beam:\begin{enumerate}
\item directly via $\bar{\nu}_e\to\bar{\nu}_\mu$ oscillations\par
$P_{\bar{\nu}_e\to\bar{\nu}_\mu}\times\Phi_{\bar{\nu}_e}$
\item via $\tau$ decays after $\bar{\nu}_e\to\bar{\nu}_\tau$ oscillations.\par
$P_{\bar{\nu}_e\to\bar{\nu}_\tau}\times BR(\tau\to\mu)\times\Phi_{\bar{\nu}_e}$
\end{enumerate}
Opposite sign muons could be produced in decays of hadrons
in neutral current events. We note however that these muons
will be soft and not isolated from the jet and can therefore
be suppressed by a mild isolation and momentum cut.

For reasons opposite to those already explained in the discussion of
events with leading electrons, the production of events opposite-sign 
muons is
suppressed in beams from $\mu^-$ decays, and enhanced in beams from 
$\mu^+$ decays. Since this enhancement is due to the presence of matter,
the energy spectrum of these events has a peak close to the resonance
energy (about 11 GeV for $\Delta m^2=3\times 10^{-3} eV^2$).\par 
The number of events for both same-sign and wrong-sign muons are shown in 
Tables \ref{tab:muo-} and \ref{tab:muo+}.
\begin{table}[tbh]
\begin{center}
\begin{tabular}{|c|c|c|c|c|c|}\hline
$E_\mu$ (GeV)&$\Delta m^2 (eV^2)$&$\bar{\nu}_e\to\bar{\nu}_\mu$&$\bar{\nu}_e\to\bar{\nu}_\tau$& unoscillated beam &$\nu_\mu\to\nu_\tau$\\ 
& &  $\mu^+$ & $\mu^+$ & $\mu^-$ & $\mu^-$ \\
\hline
  10.&                  &   1.&    0.&  315.&   17\\
  20.&$1.\times 10^{-3}$&    1.&    0.& 6096.&  115\\
  30.&                  &    3.&    0.&24342.&  265\\ \hline
  10.&                  &    4&    0&  580&    6\\
  20.&$3.\times 10^{-3}$&   22&    1& 1049&  370\\
  30.&                  &   24&    1& 7893& 1390\\ \hline
  10.&                  &    8&    0&  614&   11\\
  20.&$1.\times 10^{-2}$&   49&    2& 2784&  305\\
  30.&                  &   98&    6&14200&  555\\ \hline
\end{tabular}
\end{center}
\caption{Different contributions to leading-muon events from $\mu^-$ beams}
\label{tab:muo-}
\end{table}
\begin{table}[tbh]
\begin{center}
\begin{tabular}{|c|c|c|c|c|c|}\hline
$E_\mu$ (GeV)&$\Delta m^2 (eV^2)$&$\nu_e\to\nu_\mu$&$\nu_e\to\nu_\tau$&unoscillated beam &$\bar{\nu}_\mu\to\bar{\nu}_\tau$\\ 
& &  $\mu^-$ & $\mu^-$ & $\mu^+$ & $\mu^+$ \\
\hline
  10.&                  &   71.&    1.&  133.&    9\\
  20.&$1.\times 10^{-3}$&  104.&    3.& 2667.&   54\\
  30.&                  &  120.&    5.&10824.&  120\\ \hline
  10.&                  &  186&    4&  349&    3\\
  20.&$3.\times 10^{-3}$& 1086&   44&  562&  183\\
  30.&                  & 1703&   90& 3484&  655\\ \hline
  10.&                  &   38&    1&  260&    5\\
  20.&$1.\times 10^{-2}$&  328&   15& 1268&  142\\
  30.&                  & 3828&  282& 9045&  227\\ \hline
\end{tabular}
\end{center}
\caption{Different contributions to leading-muon events from $\mu^+$ beams}
\label{tab:muo+}
\end{table}
\par
\subsection{Events with no leading leptons}
Events with no leading leptons can be produced in \begin{enumerate}
\item neutral current processes
\item hadronic $\tau$ decays\par
$P_{\nu_\mu\to\nu_\tau}\times BR(\tau\to hadrons)\times\Phi_{\nu_\mu}+
P_{\bar{\nu}_e\to\bar{\nu}_\tau}\times BR(\tau\to hadrons)\times\Phi_{\bar{\nu}_e}$\par
\end{enumerate}
The neutral current processes do not depend on the oscillations,
so there is always an excess of events in this class, due to the hadronic 
$\tau$ decays, with respect to the expectation.\par
Total number of events for these two categories is shown in Table 
\ref{tab:nolep}, for both $\mu^-$ and $\mu^+$ beams.\par
\begin{table}[tbh]
\begin{center}
\begin{tabular}{|c|c||c|c||c|c|}\hline
$E_\mu$ (GeV)&$\Delta m^2 (eV^2)$ &  \multicolumn{2}{c|}{$\mu^-$ beam}  &  \multicolumn{2}{c|}{$\mu^+$ beam} \\
& &NC&$\tau\to$ hadrons&NC&$\tau\to$ hadrons\\ \hline
  10.&                  &  196.&   62&  173.&   35\\
  20.&$1.\times 10^{-3}$& 3534.&  429& 3168.&  213\\
  30.&                  &16061.&  986&14553.&  466\\ \hline
  10.&                  &  196&   21&  173&   25\\
  20.&$3.\times 10^{-3}$& 3534& 1383& 3168&  847\\
  30.&                  &16061& 5180&14553& 2771\\ \hline
  10.&                  &  196&   40&  173&   23\\
  20.&$1.\times 10^{-2}$& 3534& 1143& 3168&  585\\
  30.&                  &16061& 2091&14553& 1896\\ \hline
\end{tabular}
\end{center}\par
\caption{Different contributions to events with no leading leptons.}
\label{tab:nolep}
\end{table}

\section{Ratios of observed versus expected events}
Given the large oscillation probabilities, the oscillation signatures
will appear strongly. We illustrate this by computing the ratios of 
observed over expected events for the four classes.
These numbers are shown in Tables~\ref{tab:ratmu-} and \ref{tab:ratmu+}
with their statistical error.
In all cases, the effect is 
significant; to increase further the electron case, it is possible 
to take the double
ratio between events detected with negative and positive muons.\par
\begin{table}[tbh]
\begin{center}
\begin{tabular}{|c|c|c|c|c|c|}\hline
$E_\mu$ (GeV)&$\Delta m^2 (eV^2)$&$\nu_e+\bar{\nu}_e/\bar{\nu}_e^{EXP.}$&$\nu_\mu/\nu_\mu^{EXP.}$
&$\bar{\nu}_\mu/\nu_\mu^{EXP.}$&$NC/NC^{EXP}$\\ \hline
10.0& &$1.167\pm0.052$&$0.288\pm0.0158$&$0.0008\pm0.0008$&$1.318\pm0.082$\\
20.0&$1.\times 10^{-3}$&$1.058\pm0.018$&$0.714\pm0.0091$&$0.0001\pm0.0001$&$1.122\pm0.018$\\
30.0& &$1.032\pm0.010$&$0.859\pm0.0055$&$0.0001\pm0.0001$&$1.061\pm0.008$\\\hline
10.0& &$1.586\pm0.061$&$0.508\pm0.0210$&$0.0036\pm0.0018$&$1.107\pm0.075$\\ 
20.0&$3.\times 10^{-3}$&$1.397\pm0.021$&$0.163\pm0.0043$&$0.0026\pm0.0005$&$1.391\pm0.020$\\ 
30.0& &$1.249\pm0.011$&$0.324\pm0.0034$&$0.0009\pm0.0002$&$1.323\pm0.009$\\ \hline 
10.0& &$1.083\pm0.050$&$0.542\pm0.0217$&$0.0067\pm0.0024$&$1.203\pm0.078$\\
20.0&$1.\times 10^{-2}$&$1.201\pm0.019$&$0.355\pm0.0064$&$0.0059\pm0.0008$&$1.324\pm0.019$\\
30.0& &$1.561\pm0.012$&$0.515\pm0.0042$&$0.0036\pm0.0004$&$1.130\pm0.008$\\
\hline
\end{tabular}
\end{center}
\caption{Ratio between the events observed in each class for
negative muon decays, and the events that would be observed without 
oscillation. Errors are statistical only}
\label{tab:ratmu-}
\end{table}

\begin{figure}[tbh]
  \begin{center}
   \includegraphics[width=16cm]{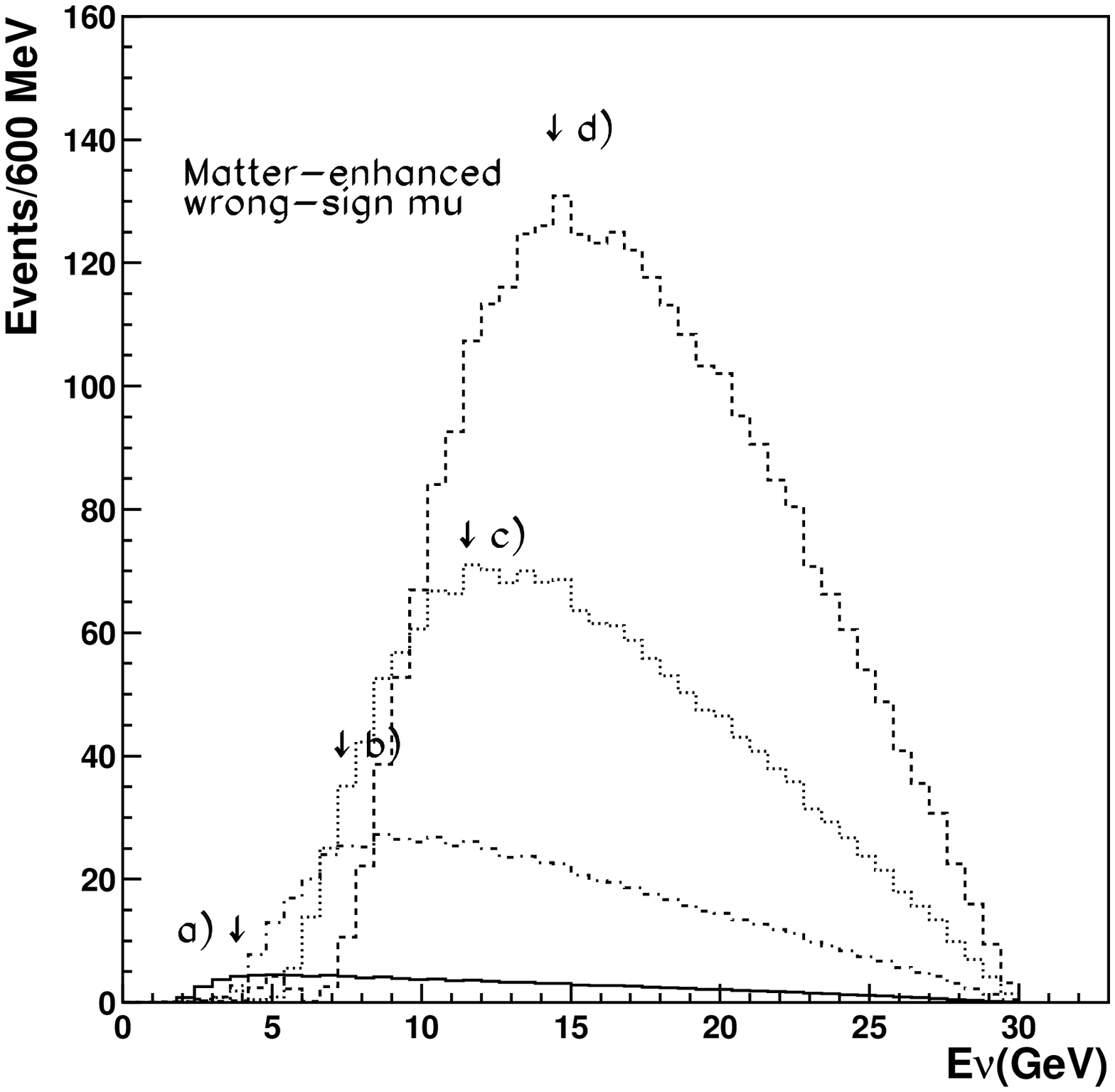}
  \end{center}
\caption{Energy spectra of events with leading $\mu^-$ in a decay beam of 
positive
muons. The different distributions refer to the following values of
$\Delta m^2 (eV^2)$: a) $1\times 10^{-3}$, b) $2\times 10^{-3}$,
c) $3\times 10^{-3}$, d) $4\times 10^{-3}$. The shift in the peak
position, corresponding to the resonance of oscillations due to
matter effects (indicated by an arrow) is clearly visible.}
\label{fig:ags3+}
\end{figure}

\begin{table}[tb]
\begin{center}
\begin{tabular}{|c|c|c|c|c|c|}\hline
$E_\mu$ (GeV)&$\Delta m^2 (eV^2)$&$\nu_e+\bar{\nu}_e/\nu_e^{EXP.}$&$\nu_\mu/\bar{\nu}_\mu^{EXP.}$
&$\bar{\nu}_\mu/\bar{\nu}_\mu^{EXP.}$&$NC/NC^{EXP}$\\ \hline
10.0& &$0.870\pm0.029$&$0.290\pm0.0244$&$0.1467\pm0.0173$&$1.202\pm0.083$\\
20.0&$1.\times 10^{-3}$&$0.980\pm0.011$&$0.714\pm0.0137$&$0.0281\pm0.0027$&$1.202\pm0.018$\\
30.0& &$0.996\pm0.006$&$0.858\pm0.0082$&$0.0098\pm0.0009$&$1.032\pm0.008$\\ \hline
10.0& &$0.646\pm0.025$&$0.720\pm0.0384$&$0.3890\pm0.0282$&$1.146\pm0.081$\\
20.0&$3.\times 10^{-3}$&$0.747\pm0.010$&$0.196\pm0.0072$&$0.2965\pm0.0088$&$1.267\pm0.020$\\
30.0& &$0.894\pm0.006$&$0.325\pm0.0050$&$0.1406\pm0.0033$&$1.190\pm0.009$\\ \hline
10.0& &$0.941\pm0.030$&$0.543\pm0.0333$&$0.0787\pm0.0127$&$1.131\pm0.081$\\
20.0&$1.\times 10^{-2}$&$0.941\pm0.011$&$0.370\pm0.0099$&$0.0900\pm0.0049$&$1.185\pm0.019$\\
30.0& &$0.718\pm0.005$&$0.727\pm0.0076$&$0.3324\pm0.0050$&$1.130\pm0.009$\\ \hline
\end{tabular}
\end{center}
\caption{Same as previous Table, but for the decay of positive 
muons.}
\label{tab:ratmu+}
\end{table}
In order to measure the oscillation parameters, it is interesting to notice
the dependence on $\Delta m^2$ of the energy spectrum of wrong-sign muons.
The position of the peak of the distribution depends on the resonance 
energy, which is a function of $\Delta m^2$.
In Figure \ref{fig:ags3+} this distribution is shown for $\Delta m^2$
values of $1,2,3, 4 \times 10^{-3}\ \rm eV^2$.\par

\section{Interpretation of experimental results}
A fit to the observed energy distributions and rates
allows the determination of the parameters $\Delta m^2$, $\theta$ and
$\phi$ governing the oscillation; the fit probability is a test of the
hypothesis that one mass scale is relevant, and that there are no other
processes or phenomena (i.e. sterile neutrinos). Given the statistical
accuracies obtained in Tables~\ref{tab:ratmu-} and \ref{tab:ratmu+},
we foresee that these fits will give good constraints on the 
oscillation parameters and scenarios.

\section{Conclusions}

The possibility of performing a second generation
very-long baseline experiment
using an intense neutrino beam from negative and positive muons 
offers the unique opportunity to study the $\dm2$ region indicated
by the atmospheric neutrinos. At the chosen optimized $L/E$ distance,
the oscillation probabilities are maximized and will allow
a complete understanding of the oscillation parameters
governing the $\nu_\mu\to\nu_\tau$ oscillation. 
Even in the case where a simple counting experiment is
performed like the one we have assumed, the visible effects are 
significant.
A large disappearance
of muon neutrinos not compensated by the appearance of taus
--- e.g. $\nu_\mu\to\nu_s$ --- would be immediately noticed.
In addition, the unique feature of the muon decay beams providing
electron and muon neutrinos in same
quantities, allow the study matter effects in Earth and given the
distance and energies involved should reveal directly the presence
of the oscillation resonance. By a change of sign in the muon storage
ring, the matter effects can be studied by directly comparing
the oscillations of neutrinos and their anti-neutrinos.

\section*{Acknowledgments}
We are indebted to R. Palmer for enlighting us on the possible upgrade
of the AGS at BNL.

\end{document}